\documentclass[final,3p,times,12pt]{elsarticle}
\renewcommand{\vec}[1]{\boldsymbol{#1}}    
\renewcommand{\geq}{\geqslant}

  \usepackage{graphicx}

\usepackage{amsmath}
\usepackage{textcomp}

\biboptions{square,numbers,sort&compress}

\journal{\rule{0pt}{30ex}}

\begin{document}

\begin{frontmatter}



\title{Screening of the electric field in covalent crystals \\ containing point defects\tnoteref{Trans}}

\tnotetext[Trans]{Translated from Izvestiya Vysshikh Uchebnykh Zavedenii, Fizika, No. 11, pp. 41-43, November, 1984. Original article submitted October 10, 1983.}


\author[BSU]{Nikolai A. Poklonski\corref{cor}}
\ead{poklonski@bsu.by}
\cortext[cor]{Current address: Physics Department, Belarusian State University, pr. Nezavisimosti 4, Minsk 220030, Belarus}

\address[BSU]{Scientific-Research Institute of Applied Physics Problems of the Belorussian State University}


\begin{abstract}
An expression is derived within the framework of the Debye--H\"uckel approximation for the screening length of a field in a \emph{p}-type semiconductor taking into account the energy spread of immobile acceptor levels and the density of states tail of the valence band. It is shown that the screening length depends additively on the product of the carrier density and their drift mobility to diffusion coefficient ratio (for free holes in valence band and holes hopping via acceptors).
\end{abstract}


\end{frontmatter}




\bigskip\bigskip

The screening of an electrostatic field solely by free electrons and holes in covalent crystalline semiconductors taking into account the density of states ``tails'' of the conduction and valence bands was discussed in [1]. The effect of a spread in defect levels on the screening of a field by electrons (holes) hopping via immobile defects was considered in [2]. Below we present, by the example of a \emph{p}-type semiconductor, the derivation of an expression for the screening length taking into account both the energy spread of the acceptor levels and the formation of a density of states tail in the valence band. 

Let us consider a crystal having a hole density \emph{p} and acceptor concentration $N = N_0 + N_{-1}$; we shall assume that all the other defects do not, in the temperature interval considered, exchange charge either with the valence band or between themselves. The electric neutrality equation has the form $p + cN = N_{-1}$, where $c$ is the degree of compensation of the acceptors. The hole density in the valence band is given by 
\begin{equation}\label{eq:01}
   p = \int_{-\infty}^{+\infty}g_pf_p\,\text{d}x_p,
\end{equation}
where $x_p = E_p/k_\text{B}T$; $E_p$ is the hole energy; $k_\text{B}$ is the Boltzmann constant; $T$ is the absolute temperature; $f_p^{-1} = 1 + \exp(y_p - x_p)$; $y_p = E_\text{F}/k_\text{B}T$; $E_\text{F}$ is the Fermi level. The density of states of free holes (taking into account spin degeneracy and the Gaussian distribution of their energy fluctuations [1]) is given by 
\begin{equation}\label{eq:02}
   g_p = \frac{2N_\varv}{\sqrt{\pi}}\frac{1}{Z_p\sqrt{2\pi}} \int_{x_p}^{\infty} \sqrt{t - x_p}\exp\left(-\frac{t^2}{2Z_p^2}\right)\, \text{d}t,
\end{equation}
where $N_\varv = 2(2\pi m_pk_\text{B}T/(2\pi\hbar)^2)^{3/2}$; $m_p$ is the effective mass of holes; $\hbar$ is the Planck constant; \linebreak $Z_p = W_p/k_\text{B}T$, $W_p$ is the mean square fluctuation in the hole energy. The concentration of negatively charged acceptors (each acceptor may capture not more than one electron per level) is 
\begin{equation}\label{eq:03}
   N_{-1} = N\cdot\overline{f}_{-1} = N \int_{-\infty}^{+\infty} g_\text{a}f_{-1}\,\text{d}x_\text{a},
\end{equation}
where the density of states $Ng_\text{a}$ and the hole occupation probability of an acceptor $f_0 = 1 - f_{-1}$ are given by 
\begin{equation}\label{eq:04}
\begin{split}
   g_\text{a} &= \frac{1}{Z_\text{a}\sqrt{2\pi}} \exp\left(-\frac{x_\text{a}^2}{2Z_\text{a}^2}\right),\quad
   f_{-1} = \frac{1}{1 + \exp(x_\text{a} + y_\text{a})};\\
   x_\text{a} &= (E_\text{a} - \overline{E_\text{a}})/k_\text{B}T,\quad
   Z_\text{a} = W_\text{a}/k_\text{B}T,\quad
   y_\text{a} = (\overline{E_\text{a}} - E_\text{F} + k_\text{B}T\ln\beta)/k_\text{B}T;\quad
\end{split}
\end{equation}
$\overline{E_\text{a}}$ is the mean value over the crystal of the acceptor ionization energy; $W_\text{a}^2$ is the variance of $E_\text{a}$; $\beta = \beta_0/\beta_{-1}$ is the degeneracy factor of the level. 

In such a semiconductor the potential $\varphi_{kq}$ of a static charge $kq$ satisfies the Poisson equation 
\begin{equation}\label{eq:05}
   \frac{1}{r^2}\cdot\frac{\text{d}}{\text{d}r}\left(r^2\cdot\frac{\text{d}\varphi_{kq}}{\text{d}r}\right) = -\frac{q}{\varepsilon}[k\cdot\delta(\vec{r}) + p(r) - N_{-1}(r) + cN],
\end{equation}
where $\varepsilon$ is the static dielectric constant of the semiconductor without taking into account the contribution of holes and of negatively charged acceptors; $q$ is the absolute value of the electron charge; $\delta(\vec{r})$ is the three-dimensional Dirac delta function; the hole density and the concentration of negatively charged acceptors at a distance $r$ from the charge $kq$ are given by expressions (1) and (3) taking into account the fact that $E_p(r) = E_p + q\varphi_{kq}(r)$ and $E_\text{a}(r) = E_\text{a} - q\varphi_{kq}(r)$. 

For $r$ values such that $q|\varphi_{kq}| \ll k_\text{B}T$, one can derive from (1)--(5) the linearized equation 
\begin{equation}\label{eq:06}
   \frac{1}{r^2}\cdot\frac{\text{d}}{\text{d}r}\left(r^2\frac{\text{d}\varphi_{kq}}{\text{d}r}\right) = - \frac{kq}{\varepsilon}\, \delta(\vec{r}) + \lambda^{-2}\cdot\varphi_{kq},
\end{equation}
whose solution with boundary conditions $\smash{\lim\limits_{r \to 0} r\cdot\varphi_{kq}(r)} = kq/4\pi\varepsilon$, $\smash{\lim\limits_{r \to \infty}\varphi_{kq}(r)} = 0$ has the form
\[
   \varphi_{kq}(r) = \frac{kq}{4\pi\varepsilon r}\exp\left(-\frac{r}{\lambda}\right),
\]
where the screening length of the field due to the charge is given by 
\begin{equation}\label{eq:07}
   \lambda^{-2} = \frac{q^2}{\varepsilon k_\text{B}T}\, \Biggl[\int_{-\infty}^{+\infty} g_pf_p(1 - f_p)\,\text{d}x_p + N\int_{-\infty}^{+\infty} g_\text{a}f_{-1}f_0\,\text{d}x_\text{a}\Biggr] \equiv \frac{q^2}{\varepsilon k_\text{B}T}\, \left(\frac{p}{\xi_p} + \frac{N\overline{f}_{-1}\cdot\overline{f}_0}{\xi_\text{a}}\right);
\end{equation}
owing to the exchange of holes with the valence band and between themselves by hopping (tunneling), the charged states of immobile acceptors migrate along the crystal and, therefore, participate in screening [3]. 

For $W_p, W_\text{a} \ll k_\text{B}T$, when $\overline{f}_{-1} \!\simeq\! f_{-1}$ and $\overline{f}_0 \simeq f_0$, it follows from (7) and (1)--(4) that $\xi_p = \xi_\text{a} \simeq 1$ and expression (7) becomes the Brooks equation [4, 5] \begin{equation}\label{eq:08}
   \lambda^{-2} = \frac{q^2}{\varepsilon k_\text{B}T}\, (p + Nf_{-1}f_0).
\end{equation}

Thus, in contrast to the opinion being expressed in the literature (see, for example, [6]) the condition $W_p, W_\text{a} \ll k_\text{B}T$ is the condition for the transition from one expression for the screening length to another (from (7) to (8)), and not the condition of the applicability of the linear screening theory (transition from Eq. (5) to (6)). Indeed, according to [7], in the presence of fluctuations of the electrostatic potential in the system ($W_p, W_\text{a} \gtrsim k_\text{B}T$), there is no point in solving the Poisson equation (5) without the linearization $q|\varphi_{kq}(r)| \ll k_\text{B}T$ leading to (6) since (5) is the equation for the mean value of the potential $\varphi_{kq}(r)$. 

For $W_p, W_\text{a} \gg k_\text{B}T$, it follows from (7) and (1)--(4) that the screening length is determined by the density of states at the Fermi level 
\[
   \lambda^{-2} = \frac{q^2}{\varepsilon k_\text{B}T}\, (p_\text{F} + N_\text{F}),
\]
where $p_\text{F} = g_p\,\big|_{x_p=y_p}$, $N_\text{F} = N\cdot g_\text{a}\,\big|_{x_\text{a} = y_\text{a}}$; the neutrality equation has, in this case, the form
\[
   \int_{y_p}^{+\infty}g_p\,\text{d}x_p + cN = N\int_{-\infty}^{y_\text{a}} g_\text{a}\,\text{d}x_\text{a}.
\]

Screening solely by free holes when $W_p \ll k_\text{B}T$ but the \emph{p}-type semiconductor is degenerate ($E_\text{F} < 0$, $|E_\text{F}| \gg k_\text{B}T$) is realized in the case when the acceptor levels lie deep in the valence band ($|\overline{E_\text{a}}| \gg |E_\text{F}|$, $N_{-1} = N$); from (2) one derives the density of states in the valence band of an ideal crystal $g_p = 2N_\varv\sqrt{-x_p/\pi}$ which determines the screening length 
\[
   \lambda_\text{f}^{-2} = \frac{q^2}{\varepsilon k_\text{B}T} \frac{p}{\xi_p},
\]
where 
\[
   p = N_\varv\cdot F_{1/2}(-y_p),\quad
   \xi_p = F_{1/2}(-y_p)/F_{-1/2}(-y_p) \geq 1;
\]
\[
   F_{1/2}(y_p) = \frac{2}{\sqrt{\pi}} \int_0^{\infty} \frac{\sqrt{t}\,\text{d}t}{1 + \exp(t - y_p)},\quad
   F_{-1/2}(y_p) = \text{d}F_{1/2}(y_p)/\text{d}y_p.
\]
It can be shown (see, [8, 9]) that in the case under consideration of screening by a degenerate hole gas $D_p/\mu_p = k_\text{B}T\xi_p/q$, where $D_p$ and $\mu_p$ are the diffusion coefficient of free holes and their drift mobility. 

Screening solely by holes hopping via acceptors dominates at low temperatures when $\overline{f}_{-1} = c$, and, according to [7], the screening length is given by $\lambda_\text{h}^{-2} \sim N_\text{h}/\xi_\text{a}$. The problem of the physical meaning of $N_\text{h} = Nc(1 - c)$ was formulated already in Brooks paper [4] and solved in [2], where it is shown that $N_\text{h}$ is the density of holes hopping via acceptors, while $\xi_\text{a} \geq 1$ indicates the extent of deviation of the their diffusion coefficient to drift mobility ratio from the classical value $k_\text{B}T/q$. Here we remark that on the basis of other premises, valid, to be sure, only for $c \ll 1$, there are indications in [10] that $N_\text{h}$ can be interpreted as the effective density of hopping holes.





\begin{thebibliography}{00}
\bibitem{1}
J. R. Lowney, A. H. Kahn, J. L. Blue, and C. L. Wilson, J. Appl. Phys., \textbf{52}, No. 6, 4075 (1981). 
\bibitem{2}
N. A. Poklonski and V. F. Stelmakh, Phys. Status Solidi (b), \textbf{117}, No. 1, 93 (1983). 
\bibitem{3}
N. A. Poklonski, V. F. Stelmakh, V. D. Tkachev, and S. V. Voitikov, Phys. Status Solidi (b), \textbf{88}, No. 2, K165 (1978). 
\bibitem{4}
H. Brooks, \emph{Theory of the Electrical Properties of Germanium and Silicon.} In: Advances in Electronics and Electron Physics (Ed. by L. Marton), Academic Press, N.Y. (1955), Vol. 7, pp. 85--182.
\bibitem{5}
L. M. Falicov and M. Cuevas, Phys. Rev., \textbf{164}, No. 3, 1025 (1967). 
\bibitem{6}
T. M. Burbaev, V. A. Kurbatov, and N. A. Penin, Sov. Phys. Semicond. \textbf{15}, No. 8, 861 (1981).
\bibitem{7}
L. P. Kudrin, Statistical Physics of Plasma [in Russian], Atomizdat, Moscow (1974), p. 15. 
\bibitem{8}
K. M. van Vliet and A. H. Marshak, Phys. Status Solidi (b), \textbf{78}, No. 2, 501 (1976). 
\bibitem{9}
P. T. Landsberg, Eur. J. Phys., \textbf{2}, No. 4, 213 (1981). 
\bibitem{10}
P. J. Price, IBM J. Res. Development, \textbf{2}, No. 2, 123 (1958). 
\end{thebibliography}



\end{document}